\documentclass[journal]{IEEEtran}
\usepackage{amssymb}
\usepackage{amsmath,amsfonts}
\usepackage{algorithmic}
\usepackage{algorithm}
\usepackage{setspace}
\usepackage{array}
\usepackage{textcomp}
\usepackage{stfloats}
\usepackage{tabularx}
\usepackage{caption} 
\usepackage{booktabs}
\usepackage{multirow}
\usepackage{diagbox}    
\usepackage{url}
\usepackage{verbatim}
\usepackage{stfloats}
\usepackage{enumitem}
\usepackage{graphicx}
\usepackage{epstopdf} 
\usepackage{cite}
\usepackage[hidelinks,colorlinks,allcolors=black]{hyperref}

\makeatletter
\def\bstctlcite#1{\@bsphack
  \@for\@citeb:=#1\do{%
    \edef\@citeb{\expandafter\@firstofone\@citeb}%
    \if@filesw\immediate\write\@auxout{\string\citation{\@citeb}}\fi}%
  \@esphack}
\makeatother
\usepackage{cleveref} 
\makeatletter
\renewcommand*{\eqref}[1]{%
  \hyperref[{#1}]{\textup{\tagform@{\ref*{#1}}}}%
}
\begin{document}

\bstctlcite{IEEEexample:BSTcontrol}

\setlength{\textfloatsep}{3pt plus 3pt minus 2pt}

\setlength{\floatsep}{5pt plus 3pt minus 2pt}

\setlength{\intextsep}{3pt plus 3pt minus 2pt}

\title{Pinching Antenna-Assisted Full-Duplex Communication Systems}

\author{Xuan Li, Xianfu Lei, Mingjiang Wu, Sotiris A. Tegos, \\ Panagiotis D. Diamantoulakis, and George K. Karagiannidis
\vspace*{-0.15 in}

\thanks{Xuan Li, Xianfu Lei are with the School of Information Science and Technology, Southwest Jiaotong University, Chengdu 610031, China (e-mails: xuanli@my.swjtu.edu.cn; xflei@swjtu.edu.cn).}
\thanks{Mingjiang Wu is with the Department of Electronic and Electrical Engineering, Southern University of Science and Technology, Shenzhen 518055, China (e-mail: wumj@sustech.edu.cn).}
\thanks{Sotiris A. Tegos, Panagiotis D. Diamantoulakis, and George K. Karagiannidis are with the Department of Electrical and Computer Engineering, Aristotle University of Thessaloniki, 54124 Thessaloniki, Greece (e-mails: tegosoti@auth.gr, padiaman@auth.gr, geokarag@auth.gr).}}

\IEEEpubid{   }

\maketitle
\begin{spacing}{0.97}
\begin{abstract}
Full-duplex (FD) communication theoretically doubles spectral efficiency. Despite this potential, its practical performance is primarily constrained by severe self-interference~(SI) and co-channel interference. Moreover, conventional fixed antenna arrays suffer from limited spatial flexibility, resulting in insufficient spatial isolation for SI suppression. To address this issue, this letter proposes an FD~architecture~assisted by pinching antenna systems (PASS). By dynamically adjusting transmit and receive antenna positions, the system enables large-scale channel reconfiguration, thereby synergizing with the base station (BS) beamforming to suppress SI and enhance the desired signal reception. To demonstrate the potential of PASS for FD~communication, we formulate a weighted sum-rate maximization problem~that jointly optimizes antenna positions, BS beamforming,~and power allocation. To tackle this non-convex problem, we~reformulate it using the weighted minimum mean square error (WMMSE) framework and develop an efficient alternating optimization (AO) algorithm to iteratively update the optimization variables. Simulation results reveal that the proposed PASS-assisted FD architecture significantly outperforms~conventional fixed antenna~arrays, achieving substantial sum-rate gains while~effectively mitigating SI to enable FD operation.\looseness=-1

\end{abstract}
\end{spacing}

\vspace{-8pt}
\section{Introduction}
\begin{spacing}{0.96} 
\IEEEPARstart{F}{ull-duplex} (FD) communication enables~simultaneous transmission and reception within the same frequency band. Theoretically, this paradigm can double spectral efficiency and significantly reduce latency, thereby supporting higher data rates and diverse services under scarce spectrum resources \cite{ref1}. However, practical FD implementation is hindered by self-interference (SI) at~the base station~(BS), as well as co-channel interference (CCI) between uplink and downlink users induced by resource multiplexing.~In particular,~severe SI induced by antenna leakage, spatial coupling, and hardware nonlinearities is the primary bottleneck, as its power typically exceeds that of the desired signal~by more than 100 dB \cite{ref2}, which potentially overwhelms the intended reception in the absence of effective suppression. Current SI suppression techniques generally employ a cascade architecture comprising spatial, analog, and digital domains \cite{ref3}. To prevent the receiver front-end saturation induced by high-power SI, it is necessary to suppress SI to a level~within~the receiver's dynamic range in the spatial and analog domains \cite{ref1}. However, existing spatial suppression schemes are mainly based on fixed antenna arrays with constrained flexibility, resulting in insufficient intrinsic spatial isolation and increasing the burden on subsequent analog SI cancellation implemented with~digital controllers and~passive elements (e.g., attenuators, phase shifters, delay lines) \cite{ref1}. Unfortunately, designing such hardware and parameters incurs additional costs, complexity, and limited scalability. Moreover, inherent hardware impairments compromise the precision \cite{ref4}.
\setlength{\parskip}{0pt}

To address the physical limitations of conventional~fixed antenna arrays, the pinching antenna (PA), proposed~by NTT DOCOMO in 2022 as a new paradigm~for~reconfigurable antennas \cite{ref5}, offers new opportunities to overcome~the~aforementioned spatial~isolation bottleneck in FD systems. By enabling large-scale adjustment of antenna positions along dielectric waveguides spanning tens of meters, pinching antenna systems (PASS) facilitate both large- and small-scale reconfiguration of wireless channel characteristics at minimal hardware cost \cite{ref6}. Recent studies have investigated the throughput advantages of PASS in unidirectional~uplink~and downlink scenarios, demonstrating its considerable potential in half-duplex communications \cite{ref7},~\cite{ref8},~\cite{ref9},~\cite{ref10}. Nevertheless, the application of PASS in FD systems remains in~its early~stages. The joint uplink and downlink optimization in PASS was initially investigated in~\cite{ref11}, providing valuable insights into its overall sum-rate advantages over~time-division duplex systems. Although the authors in~\cite{ref11}~investigated antenna
activation for PASS-assisted FD systems via a spatial-correlation-based algorithm, how to fully exploit the unique channel reconfiguration capability of PASS and synergistically integrate it with BS beamforming to mitigate critical FD bottlenecks remains an open problem. \looseness=-1

Motivated by the above considerations, this letter proposes a PASS-assisted FD architecture. Specifically, depending on the users' geometry positions, dynamic antenna placement~optimizes the spatial channel to effectively mitigate large-scale path loss, which ultimately enhances the reception of desired signals and increases the tolerance of the system to residual SI. Meanwhile, the passive spatial isolation provided by optimized relative positions of transmit and receive PAs synergizes with active BS beamforming to establish a joint interference suppression mechanism. The considered objective is to maximize the weighted sum-rate of the system by~jointly optimizing the PA positions, BS beamforming, and user transmit power. To tackle the formulated non-convex problem, we first reformulate it utilizing the weighted minimum mean square error (WMMSE) method. Subsequently, an alternating optimization (AO) algorithm is developed to iteratively solve the variables, yielding closed-form optimal solutions for the BS beamforming and uplink user transmit power. In addition, PA positions are sequentially optimized using an element-wise one-dimensional (1D) search algorithm. Finally, simulation results validate the performance advantages of the proposed PASS-assisted FD communication systems.
\end{spacing}
\begin{spacing}{0.975} 
\section{System Model And Problem Formulation}
\begin{figure}[!t]
\vspace{-0.8em}
\centering
\includegraphics[width=3.27in]{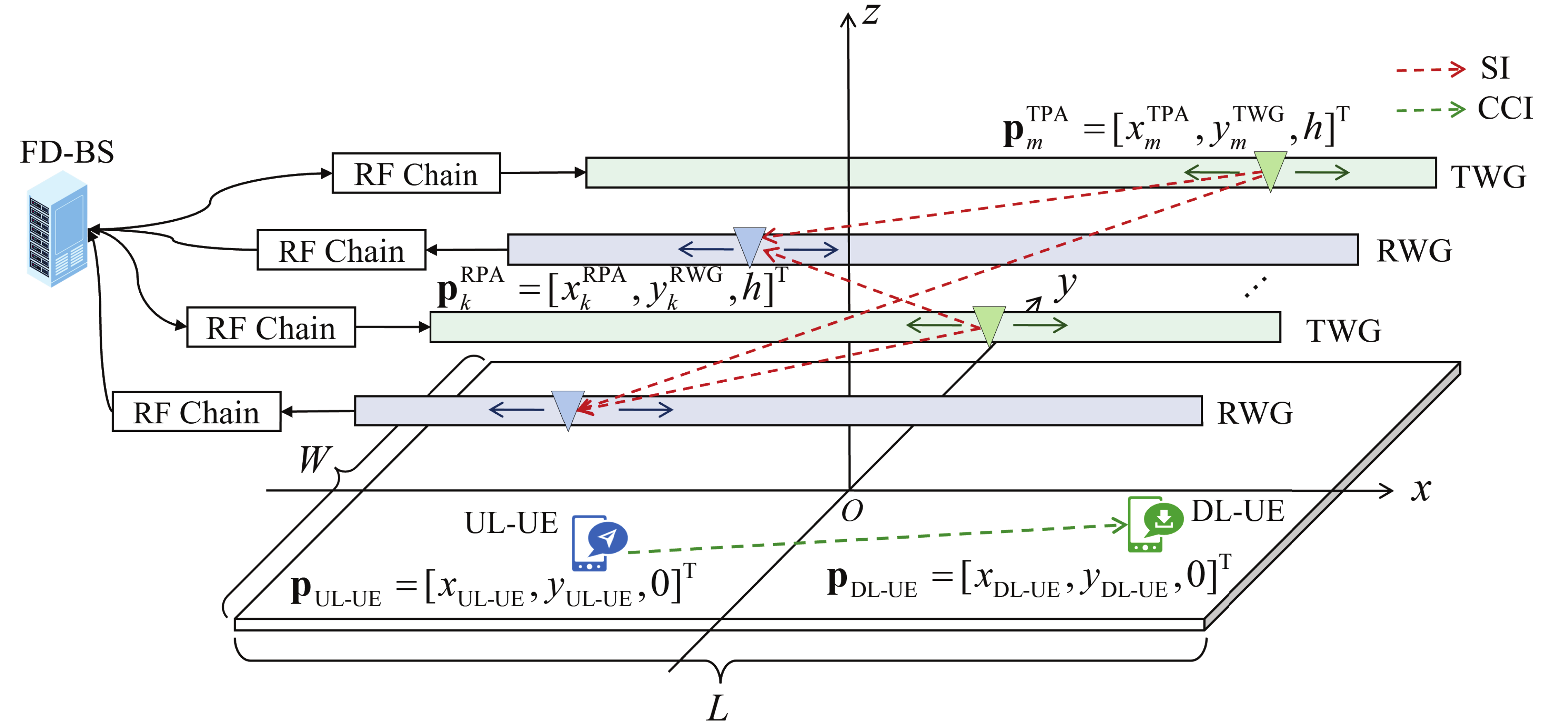}
\captionsetup{font=scriptsize, skip=2pt} 
\captionsetup{justification=raggedright, singlelinecheck=false}
\caption{System model for PASS-assisted FD communication.}
\label{fig_1}
\end{figure}

%\vspace*{-0.2cm} 
We consider a PASS-assisted FD communication system as illustrated in Fig. \ref{fig_1}. 
The system comprises an FD-BS featuring $M$ transmit waveguides (TWGs) and $K$ receive waveguides (RWGs). Each waveguide is fed by a dedicated radio frequency chain, and for analytical tractability and system symmetry, we assume one transmit pinching antenna (TPA) per TWG and one receive pinching antenna (RPA) per RWG. Multi-PA-per-waveguide configurations, which introduce the \textit{inter-antenna radiation} effect in the uplink, are left for future work~\cite{ref12}. Let $\mathcal{M} \triangleq \{1,\dots,M\}$ and $\mathcal{K} \triangleq \{1,\dots,K\}$ denote the index sets of TPAs and RPAs, respectively. Assuming that all waveguides have an identical length of $L$, they are deployed in parallel and  interleaved with uniform spacing along the width $W$ of the service region. We establish a three-dimensional Cartesian coordinate system with its origin at the center of the service region, where all waveguides are parallel to the $x$-axis. 
The coordinates of the $m$-th TPA and $k$-th RPA are denoted by $\boldsymbol{\mathbf{p}}_m^{\text{TPA}} = [x_m^{\text{TPA}}, y_m^{\text{TWG}}, h]^\text{T}$ and $\boldsymbol{\mathbf{p}}_k^{\text{RPA}} = [x_k^{\text{RPA}}, y_k^{\text{RWG}}, h]^\text{T}$, respectively. The height of the waveguide $h$ and $y$-axis positions $\{y_m^{\text{TWG}}, y_k^{\text{RWG}}\}$ are pre-determined by the waveguide configuration. The system provides simultaneous communication services to a single-antenna downlink user (DL-UE) and a single-antenna uplink user (UL-UE), and the positions of the DL-UE and UL-UE are given by $\boldsymbol{\mathbf{p}}_{\text{DL-UE}} = [x_{\text{DL-UE}}, y_{\text{DL-UE}}, 0]^\text{T}$ and $\boldsymbol{\mathbf{p}}_{\text{UL-UE}} = [x_{\text{UL-UE}}, y_{\text{UL-UE}}, 0]^\text{T}$, respectively.
\vspace*{-0.15cm}
\subsection{System Model}
\textit{1) Downlink Channel}: According to the classical spherical wave channel model, the downlink channel from the $m$-th TPA to the DL-UE can be expressed as
\begin{equation}
\label{eq:1}
h_{\text{DL},m} = \frac{\sqrt{\eta} e^{-j 2\pi \left( \frac{\|\mathbf{p}_{\text{DL-UE}} - \mathbf{p}_m^{\text{TPA}}\|}{\lambda} + \frac{\|\mathbf{p}_{m,0}^{\text{TWG}} - \mathbf{p}_m^{\text{TPA}}\|}{\lambda_g} \right)}}{\|\mathbf{p}_{\text{DL-UE}} - \mathbf{p}_m^{\text{TPA}}\|},
\end{equation}
where $\eta\! = \!c^2 \!/ \!(4\pi f_c)^2$, $c$ is the speed of light, $f_c$ is the~carrier frequency, $\lambda \!=\! c/f_c$ is the wavelength of the signal in~free space, $\lambda_g \!= \!c/(f_c n_{\text{eff}})$ is the wavelength of the signal inside the waveguide, and $n_{\text{eff}}$ is the effective refractive index of the dielectric waveguide. The term $\frac{2\pi}\lambda\|\mathbf{p}_{\text{DL-UE}} \!-\! \mathbf{p}_m^{\text{TPA}}\|$ represents the phase shift introduced by signal propagation from the $m$-th TPA to the DL-UE in free space, while $\frac{2\pi} {\lambda_g}\|\mathbf{p}_{m,0}^{\text{TWG}} \!- \!\mathbf{p}_m^{\text{TPA}}\| $ represents the phase shift introduced by signal propagation inside the $m$-th TWG. Additionally, $\mathbf{p}_{m,0}^{\text{TWG}} \!= \![-L/2, y_m^{\text{TWG}}, h]^\text{T}$ represents the position of the feed point of the $m$-th TWG.

\textit{2) Uplink Channel}: The channel from the UL-UE to the $k$-th RPA can be expressed as
\begin{equation}
h_{\text{UL},k} = \frac{\sqrt{\eta} e^{-j 2\pi \left( \frac{\|\mathbf{p}_k^{\text{RPA}} - \mathbf{p}_{\text{UL-UE}}\|}{\lambda} + \frac{\|\mathbf{p}_{k,0}^{\text{RWG}} - \mathbf{p}_k^{\text{RPA}}\|}{\lambda_g} \right)}}{\|\mathbf{p}_k^{\text{RPA}} - \mathbf{p}_{\text{UL-UE}}\|},
\end{equation}
where $\frac{2\pi}\lambda\|\mathbf{p}_{\text{UL-UE}}\! -\! \mathbf{p}_k^{\text{RPA}}\|$ and $\frac{2\pi}{ \lambda_g}\|\mathbf{p}_{k,0}^{\text{RWG}} \!-\! \mathbf{p}_k^{\text{RPA}} \|$ represent the phase shifts introduced by the free-space and RWG propagations, respectively, and $\mathbf{p}_{k,0}^{\text{RWG}} \!=\! [-L/2, y_k^{\text{RWG}}, h]^\text{T}$ is the feed point of the $k$-th RWG.

\textit{3) Interference Channel}: Under the FD communication mode, due to simultaneous uplink and downlink transmissions in the same frequency band, the RPAs will be subjected to SI from the TPAs. Consequently, the SI channel between the $m$-th TPA and the $k$-th RPA can be modeled as
\begin{equation}
h_{\text{SI},k,m} = \frac{\sqrt{\eta} e^{-j 2\pi \left( \frac{\|\mathbf{p}_k^{\text{RPA}} - \mathbf{p}_m^{\text{TPA}}\|}{\lambda} + \theta_{k,m} \right)}}{\|\mathbf{p}_k^{\text{RPA}} - \mathbf{p}_m^{\text{TPA}}\|},
\end{equation}
where $\frac{2\pi}{\lambda}{\|\mathbf{p}_k^{\text{RPA}} - \mathbf{p}_m^{\text{TPA}}\|}$ captures the free-space phase shift, while $2\pi\theta_{k,m} = \frac{2\pi}{\lambda_g}{\|\boldsymbol{\mathbf{p}}_{m,0}^{\text{TWG}} - \boldsymbol{\mathbf{\mathbf{p}}}_m^{\text{TPA}}\|} + \frac{2\pi}{\lambda_g}{\|\boldsymbol{\mathbf{p}}_{k,0}^{\text{RWG}} - \boldsymbol{\mathbf{p}}_k^{\text{RPA}}\|}$ characterizes the phase shift inside the waveguides. Subsequently, the equivalent SI channel $\mathbf{H}_{\text{SI}}$ can be expressed as
\begin{equation}
\mathbf{H}_{\text{SI}} = \sqrt{\frac{1}\rho} \begin{bmatrix}
h_{\text{SI},1,1} & \cdots & h_{\text{SI},1,M} \\
\vdots & \ddots & \vdots \\
h_{\text{SI},K,1} & \cdots & h_{\text{SI},K,M}
\end{bmatrix}\in \mathbb{C}^{K \times M} ,
\end{equation}
where $\rho$ denotes the effective interference cancellation depth achieved by the analog SI cancellation technology \cite{ref13}. Additionally, the CCI channel between half-duplex users is $h_{\text{CCI}} \sim \mathcal{CN}(0, \varsigma(d_{\text{CCI}}))$, where $\varsigma(\cdot)$ denotes the large-scale path loss determined by the distance $d_{\text{CCI}}$ between users.
%\setlength{\parskip}{0pt}

%\vspace*{-0.08cm} 
Let $\mathbf{s}_{\text{DL}}\! =\! \mathbf{w}s_{\text{DL}} \!\in \!\mathbb{C}^{M \times 1}$ denote the downlink transmitted signal vector, where $\mathbf{w} \!\in\! \mathbb{C}^{M \times 1}$ is the beamforming vector and $s_{\text{DL}} \!\in\! \mathbb{C}$ is the data symbol. Similarly, let $s_{\text{UL}} \!\in\! \mathbb{C}$~be the uplink data symbol. Assume $s_{\text{DL}}$, $s_{\text{UL}}$ are statistically independent of each other, and both of them satisfy $\mathbb{E}\{|s_{\text{DL}}|^2\} \!=\! \mathbb{E}\{|s_{\text{UL}}|^2\}\!=\!1$. Consequently, the signal~received at the DL-UE is given by
\begin{equation}
\label{eq:downlink_signal} 
y_{\text{DL}} = {\mathbf{h}_{\text{DL}}^\text{H} \mathbf{s}_{\text{DL}}}+ {\sqrt{p_\text{t}} h_{\text{CCI}} s_{\text{UL}}}+ {n_{\text{DL-UE}},}
\end{equation}
where $\mathbf{h}_{\text{DL}}\triangleq [h_{\text{DL},1}, \dots, h_{\text{DL},M}]^\text{T}\in \mathbb{C}^{M \times 1}$ represents the downlink channel vector from all TPAs to the DL-UE, $p_\text{t}$ is the transmit power of the UL-UE, and $n_{\text{DL-UE}} \sim \mathcal{CN}(0, \sigma_{\text{DL-UE}}^2)$ is the additive white gaussian noise (AWGN) at the DL-UE. Therefore, the received signal-to-interference-plus-noise ratio (SINR) at the DL-UE is expressed as
\begin{equation}
\Gamma_{\text{DL}} = \frac{|\mathbf{h}_{\text{DL}}^\text{H} \mathbf{w}|^2}{|h_{\text{CCI}}|^2 p_\text{t} + \sigma_{\text{DL-UE}}^2}.
\end{equation}

Regarding the uplink, the uplink signal vector $\mathbf{y}_{\text{UL}} \in \mathbb{C}^{K \times 1}$ received by the FD-BS is given by
\begin{equation}
\label{eq:uplink_signal} 
\mathbf{y}_{\text{UL}} = {\sqrt{p_\text{t}} \mathbf{h} _{\text{UL}}s_{\text{UL}}} + {\mathbf{H}_{\text{SI}} \mathbf{s}_{\text{DL}}} + {\mathbf{n}_{\text{FD-BS}},}
\end{equation}
where $\mathbf{h}_{\text{UL}}\triangleq [h_{\text{UL},1}, \dots, h_{\text{UL},K}]^\text{T}\in \mathbb{C}^{K\times 1}$ represents the uplink channel vector from the UL-UE to all RPAs, $\mathbf{n}_{\text{FD-BS}} = [n_1, \dots, n_K]^\text{T} \in \mathbb{C}^{K \times 1}$ is the AWGN vector at the FD-BS, satisfying $\mathbf{n}_{\text{FD-BS}} \sim \mathcal{CN}(\mathbf{0}_K, \sigma_{\text{FD-BS}}^2 \mathbf{I}_K)$. Assuming that the FD-BS employs a receiver combining vector $\mathbf{v} \!\in\! \mathbb{C}^{K \times 1}$ for the uplink signal combining, the SINR at the FD-BS is expressed as \looseness=-1
\begin{equation}
\label{eq:8} 
\Gamma_{\text{UL}} = \frac{p_\text{t} |\mathbf{v}^\text{H} \mathbf{h}_{\text{UL}}|^2}{|\mathbf{v}^\text{H} \mathbf{H}_{\text{SI}} \mathbf{w}|^2 + \sigma_{\text{FD-BS}}^2 \|\mathbf{v}\|^2},
\end{equation} 
where $|\mathbf{v}^\text{H} \mathbf{H}_{\text{SI}} \mathbf{w}|^2$ is the residual SI power. By jointly adjusting the spatial channel and spatial filtering, we can suppress the residual SI while preserving the desired signal gains.

\textit{Remark 1}: In FD systems, the SI channel $\mathbf{H}_{\text{SI}}$ usually~has~a high channel gain. Due to the limited dynamic range of practical receivers \cite{ref1}, \cite{ref2}, \cite{ref14}, the strong SI signal ${\mathbf{H}_{\text{SI}} \mathbf{s}_{\text{DL}}}$ may introduce additional noise and distortion at the receiver side. Therefore, suppressing the power of the SI signal in~the spatial-domain is crucial for FD systems, as evaluated in~Section IV.

\vspace*{-0.3cm}
\subsection{Problem Formulation}
We aim to maximize the weighted sum-rate of the system by jointly optimizing the antenna positions $\mathbf{x}^{\text{TPA}} \triangleq [x_1^{\text{TPA}}, \dots, x_M^{\text{TPA}}]^\text{T}$ and $\mathbf{x}^{\text{RPA}} \triangleq [x_1^{\text{RPA}}, \dots, x_K^{\text{RPA}}]^\text{T}$, the FD-BS transceiver beamforming vectors $\mathbf{w}, \mathbf{v}$, and the UL-UE transmit power $p_\text{t}$. This optimization problem is formulated as: \looseness=-1
\begin{subequations}
\begin{alignat}{2}
\label{eq:9a} 
& \textbf{P1}: \!\!\max_{\substack{\mathbf{x}^\text{RPA}, \mathbf{x}^\text{TPA} \\ \mathbf{w}, \mathbf{v}, p_\text{t}}} \quad &&\!\! \omega_{\text{DL}}\log_2(1 + \Gamma_{\text{DL}}) + \omega_{\text{UL}}\log_2(1 + \Gamma_{\text{UL}}) \tag{9a} \\
\label{eq:9b} 
& \ \ \ \ \ \ \ \ \ \text{s.t.} \quad && \!\!\|\mathbf{w}\|^2 \leq P_{\text{BS}}, \tag{9b} \\
\label{eq:9c} 
& && \!\!0 \leq p_\text{t} \leq P_\text{t}, \tag{9c} \\
\label{eq:9d} 
& && \!\!\!-\frac{L}{2} \leq x_m^{\text{TPA}} \leq \frac{L}{2}, \forall m \in \mathcal{M}, \tag{9d} \\
\label{eq:9e} 
& && \!\!\!-\frac{L}{2} \leq x_k^{\text{RPA}} \leq \frac{L}{2}, \forall k \in \mathcal{K}, \tag{9e}
\end{alignat}
\end{subequations}
where constraints \eqref{eq:9b} and \eqref{eq:9c} limit the maximum transmit power for the FD-BS and the UL-UE, respectively. Constraints \eqref{eq:9d} and \eqref{eq:9e} specify the position constraints for the TPAs and RPAs, respectively, ensuring that the antenna positions remain within the physical range of the waveguides. Moreover, $\omega_{\text{DL}}$ and $\omega_{\text{UL}}$ are the weights of DL-UE and UL-UE, respectively.

Problem \textbf{P1} is a challenging non-convex problem, whose complexity primarily stems from two aspects: i) the fractional-logarithmic coupling in the objective function; and ii) the embedding of the antenna positions in the highly non-linear complex exponential components of the channel responses. To tackle the intractability, we first employ the WMMSE method to transform the objective into an equivalent, more tractable weighted MSE. Subsequently, an AO framework is developed to iteratively solve the reformulated problem.
\vspace*{-0.2cm}
\section{Problem Reformulation And Algorithm Design}
%\vspace*{-0.05cm}

Following the WMMSE framework \cite{ref15}, we introduce the receiver equalization coefficient $u \!\in \!\mathbb{C}$ for DL-UE, the MSEs of DL and UL signals are derived from \eqref{eq:downlink_signal} and \eqref{eq:uplink_signal} as follows:
\begin{align}
\varepsilon_{\text{DL}} &= \mathbb{E} \left\{ \left| s_{\text{DL}} - u^* y_{\text{DL}} \right|^2 \right\} \notag \\
&= |1 - u^* \mathbf{h}_{\text{DL}}^\text{H} \mathbf{w}|^2 + |u|^2 (|h_{\text{CCI}}|^2 p_\text{t} + \sigma_{\text{DL-UE}}^2), \label{eq:10} \\
\varepsilon_{\text{UL}} &= \mathbb{E} \left\{ \left| s_{\text{UL}} - \mathbf{v}^{\text{H}} \mathbf{y}_{\text{UL}} \right|^2 \right\} \notag \\
&= |1 - \sqrt{p_\text{t}} \mathbf{v}^\text{H} \mathbf{h}_{\text{UL}}|^2 + |\mathbf{v}^\text{H} \mathbf{H}_{\text{SI}} \mathbf{w}|^2 + \sigma_{\text{FD-BS}}^2 \|\mathbf{v}\|^2, \label{eq:11}
\end{align}
where $($·$)^*$ represents the conjugate operation.

Subsequently,~by leveraging the similar idea as in \cite[Theorem 1]{ref15} and introducing the auxiliary MSE weights $\alpha, \beta > 0$, problem \textbf{P1} can be transformed into the following form, which shares the equivalent optimal solutions $\{\mathbf{v}, \mathbf{w}, p_t, \mathbf{x}^{\mathrm{RPA}}, \mathbf{x}^{\mathrm{TPA}}\}$ with the original formulation:
\begin{subequations}
\begin{alignat}{2}
\label{eq:12a}
\textbf{P2}: \!\!\!\!\ & \min_{\substack{\mathbf{x}^\text{RPA}, \mathbf{x}^\text{TPA}, u \\ \mathbf{v}, \mathbf{w}, p_\text{t},\alpha, \beta}} \quad && \!\!\!\!\alpha \varepsilon_{\text{DL}} \! - \!\omega_{\text{DL}}\log_2(\frac{\text{ln2}\alpha}{\omega_{\text{DL}}}) \! + \!\beta \varepsilon_{\text{UL}} \!- \!\omega_{\text{UL}}\log_2(\frac{\text{ln2}\beta}{\omega_{\text{UL}}}) \tag{12} \\
&\ \ \ \ \ \text{s.t.} \quad && \!\!\!\!\!\eqref{eq:9b}, \eqref{eq:9c}, \eqref{eq:9d}, \eqref{eq:9e}. \nonumber
\end{alignat}
\end{subequations}

While the objective function of \textbf{P2} eliminates the intractable fractional-logarithmic form of \eqref{eq:9a}, the optimization variables are still highly coupled, motivating us to solve them alternately via an AO framework.

\textit{1) Subproblem with respect to MSE Weights $\alpha$ and $\beta$:} For given $\{\mathbf{v}, u, p_t, \mathbf{w}, \mathbf{x}^{\text{RPA}}, \mathbf{x}^{\text{TPA}}\}$, the objective function of \textbf{P2} is strictly convex with respect to the MSE weights $\alpha$ and $\beta$. Therefore, by applying the first-order optimality conditions, the optimal closed-form updates for $\alpha$ and $\beta$ are obtained as
\begin{equation}
\label{eq:13}
\alpha_{\text{opt}} = \frac{\omega_{\text{DL}}}{\ln 2} \left( \varepsilon_{\text{DL}} \right)^{-1},
\beta_{\text{opt}} = \frac{\omega_{\text{UL}}}{\ln 2} \left( \varepsilon_{\text{UL}} \right)^{-1}.
\end{equation}

\textit{2) Subproblem with respect to Receiver Design for $\mathbf{v}$ and~$u$:}
For given $\{\alpha,\ \beta, p_t, \mathbf{w}, \mathbf{x}^{\text{RPA}}, \mathbf{x}^{\mathrm{TPA}}\}$, \textbf{P2} can be decoupled into two unconstrained convex quadratic problems with respect to the FD-BS receiver combining vector $\mathbf{v}$ and the DL-UE equalization coefficient $u$. By setting their respective first-order derivatives to zero, we obtain the MMSE receivers as follows:
\begin{align}
\mathbf{v}_{\text{opt}} & =  \frac{\sqrt{p_\text{t}}}{\sigma_{\text{FD-BS}}^2}\left( \mathbf{I}_{K}\!-\!\mathbf{C}\left(\sigma_{\text{FD-BS}}^2\mathbf{I}_{2} \!+\! \mathbf{C}^\text{H}\mathbf{C}\right)^{-1}\mathbf{C}^\text{H}\right)\mathbf{h}_{\text{UL}}, \label{eq:14}
\\
u_{\text{opt}} &= \frac{\mathbf{h}_{\text{DL}}^\text{H} \mathbf{w}}{|\mathbf{h}_{\text{DL}}^\text{H} \mathbf{w}|^2 + |h_{\text{CCI}}|^2 p_\text{t} + \sigma_{\text{DL-UE}}^2}, \label{eq:15}
\end{align}
where \eqref{eq:14} follows from the Woodbury matrix identity, and $\mathbf{C} \triangleq \begin{bmatrix} \sqrt{p_t} \mathbf{h}_{\mathrm{UL}},  \mathbf{H}_{\mathrm{SI}} \mathbf{w} \end{bmatrix} \in \mathbb{C}^{K \times 2}$.

\textit{3) Subproblem with respect to \text{UL-UE} Transmit Power $p_t$:} For given $\{\alpha, \beta,\mathbf{v}, u, \mathbf{w}, \mathbf{x}^{\text{RPA}}, \mathbf{x}^{\mathrm{TPA}}\}$, the UL-UE transmit power optimization problem is formulated as
\begin{subequations}
\begin{alignat}{2}
\label{eq:16}
\min_{p_\text{t}} \quad & \left( \alpha |u|^2 |h_{\text{CCI}}|^2 + \beta | \mathbf{v}^\text{H} \mathbf{h}_{\text{UL}} |^2 \right) p_\text{t} - 2\sqrt{p_\text{t}} \beta \mathcal{R} \{ \mathbf{v}^\text{H} \mathbf{h}_{\text{UL}} \} \tag{16} \\[-10pt]
\text{s.t.} \quad & \eqref{eq:9c}, \nonumber
\end{alignat}
\end{subequations}
where $\mathcal{R}\{\cdot\}$ denotes the real-part operator. 

Introducing~the auxiliary variable $q = \sqrt{p_\text{t}}$, the objective function in \eqref{eq:16} is transformed into a standard convex quadratic form with~respect to $q$. Then, considering the unconstrained case, the optimal solution for $q$ is obtained via the first-order optimality condition as
\begin{equation}
\label{eq:17}
q_{\text{opt}} = \frac{\beta \mathcal{R} \{ \mathbf{v}^\text{H} \mathbf{h}_{\text{UL}} \}}{\alpha |u|^2 |h_{\text{CCI}}|^2 + \beta | \mathbf{v}^\text{H} \mathbf{h}_{\text{UL}} |^2}.
\end{equation}

Subsequently, considering the practical transmit power limit in \eqref{eq:9c}, the optimal closed-form solution for $p_\text{t}$ is obtained by projecting $q_{\text{opt}}$ onto the feasible region $[0, \sqrt{P_\text{t}}]$:
\begin{equation}
\label{eq:18}
p_{\text{t},\text{opt}} = \left( \max \left( 0, \min \left( \sqrt{P_\text{t}}, q_{\text{opt}} \right) \right) \right)^2.
\end{equation}

\textit{4) Subproblem with respect to the Downlink Transmit Beamforming Vector $\mathbf{w}$:} For given $\{\alpha, \beta,\mathbf{v}, u, p_t, \mathbf{x}^{\text{RPA}}, \mathbf{x}^{\mathrm{TPA}}\}$, the problem for optimizing $\mathbf{w}$ is formulated as
\begin{subequations}
\begin{alignat}{2}
\label{eq:19}
\min_{\mathbf{w}}\ \mathbf{w}^\text{H} \mathbf{A} \mathbf{w} - \mathbf{w}^\text{H} \mathbf{b} - \mathbf{b}^\text{H} \mathbf{w}, \quad \text{s.t. } \eqref{eq:9b}, \tag{19}
\end{alignat}
\end{subequations}
where $\mathbf{A} = \alpha|u|^2\mathbf{h}_{\text{DL}}\mathbf{h}_{\text{DL}}^\text{H} + \beta\mathbf{H}_{\text{SI}}^\text{H}\mathbf{v}\mathbf{v}^\text{H}\mathbf{H}_{\text{SI}} \succeq \mathbf{0}$, $\mathbf{b} = \alpha u\mathbf{h}_{\text{DL}}$. 

To solve this convex quadratically constrained quadratic program, we introduce the Lagrange multiplier $\mu \geq 0$ and construct the Lagrangian as
\begin{equation}
\label{eq:20}
\mathcal{L}(\mathbf{w}, \mu) = \mathbf{w}^\text{H} \mathbf{A} \mathbf{w} - \mathbf{w}^\text{H} \mathbf{b} - \mathbf{b}^\text{H} \mathbf{w} + \mu (\mathbf{w}^\text{H} \mathbf{w} - P_{\text{BS}}).
\end{equation}

By setting $\frac{\partial \mathcal{L}}{\partial \mathbf{w}^*} = 0$, the optimal beamformer is derived as
\begin{equation}
\label{eq:21}
\mathbf{w}_{\text{opt}} = (\mathbf{A} + \mu \mathbf{I})^{-1} \mathbf{b},
\end{equation}
where the Lagrange multiplier $\mu$ can be determined via bisection search. To further reduce computational complexity, the Hermitian matrix $\mathbf{A}$ is preprocessed by eigen value decomposition (EVD) as $\mathbf{A}=\mathbf{Q}\mathbf{\Sigma}\mathbf{Q}^{\mathrm H}$, which yields
\begin{equation}
\label{eq:22}
g(\mu)=\|(\mathbf{A}+\mu\mathbf{I})^{-1}\mathbf{b}\|^2
=\sum_{i=1}^{M}\frac{|[\tilde{\mathbf b}]_i|^2}{([\mathbf{\Sigma}]_i+\mu)^2},
\end{equation}
where $\tilde{\mathbf b}=\mathbf{Q}^{\mathrm H}\mathbf b$. Since $\tilde{\mathbf b}$ and $\mathbf\Sigma$ are fixed during the bisection search, each iteration only involves scalar computations.

\textit{5) Subproblem with respect to the Receive and Transmit Antenna Positions} $\{\mathbf{x}^{\text{RPA}}, \mathbf{x}^{\text{TPA}}\}$\textit{:} 
For given $\{\alpha, \beta,\mathbf{v}, u, \mathbf{w}, p_t\}$, the problem for jointly optimizing $\mathbf{x}^{\text{RPA}}$ and $\mathbf{x}^{\text{TPA}}$ is given~by \looseness=-1
\begin{subequations}
\label{eq:23}
\begin{alignat}{2}
\min_{\mathbf{x}^{\text{RPA}},\mathbf{x}^{\text{TPA}}} \alpha \varepsilon_{\mathrm{DL}} + \beta \varepsilon_{\mathrm{UL}}
\quad \mathrm{s.t.}\ \eqref{eq:9d},\eqref{eq:9e}.
\label{eq:prob_23}\tag{23}
\end{alignat}
\end{subequations}

The objective function in problem \eqref{eq:23} is non-convex, and due to the strong coupling between the TPA and RPA positions introduced by the SI channel, direct joint optimization is computationally intractable. To tackle this, we decouple the antenna positions into two independent blocks (TPAs and RPAs) and update them alternately.  Specifically, the optimal coordinates within each block are alternately determined via a Gauss-Seidel-based element-wise 1D search over the feasible waveguide region by fixing the remaining antennas at their latest updated positions. 

The proposed algorithm to solve \textbf{P2} is summarized in Algorithm \ref{Algorithm 1}. At each iteration step, every variable block is updated to minimize the overall objective function while fixing the remaining variables. This ensures that the objective value of \textbf{P2} is monotonically non-increasing. Given that the objective function is inherently lower-bounded, the proposed AO-based algorithm is guaranteed to converge. Additionally, the updates of $\alpha$ and $\beta$ have a complexity of $O(MK)$. The updates of $u$ and $p_t$ have complexities of $O(M)$ and $O(K)$, respectively. The complexity of updating $\mathbf v$ is $O(MK)$ by exploiting the Woodbury matrix identity. The complexity of updating $\mathbf w$ is $O(M^3 + MK + MI_\mu)$, where $O(M^3)$ arises from the EVD of $\mathbf A\in\mathbb C^{M\times M}$ and $O(MI_\mu)$ comes from the $I_\mu$ iterations of the bisection search. The element-wise 1D search for $\{\mathbf{x}^{\text{RPA}}, \mathbf{x}^{\text{TPA}}\}$ has a complexity of $O((M+K)N_xMK)$, where $N_x$ is the number of candidate points. Therefore, the overall complexity of Algorithm \ref{Algorithm 1} is $O(I_{\mathrm{AO}}(\!M^3\!+\!MK\!+\!MI_\mu\!+\!(M+K)N_xMK))$, where $I_{\mathrm{AO}}$ denotes the number of AO iterations.

\vspace*{-0.2cm} 
\section{Simulation Results}

This section provides numerical results to illustrate~the performance of the proposed PASS-assisted FD~architecture. We consider a service region with size $L \times W = 40 \!\text{ m} \times 10 \!\text{ m}$, where both uplink and downlink users are randomly located within the service region. The transmit and receive waveguides are deployed parallel to the $x$-axis with a height of $h \!=\! 3$ m, and are interleaved with uniform spacing along the width $W$ of the service region. The noise powers at the FD-BS and DL-UE are assumed to be $\sigma_{\text{FD-BS}}^2 \!= \!\sigma_{\text{DL-UE}}^2 =\! -90$ dBm~\cite{ref8}. The carrier frequency is $f_c \!=\! 28$~GHz, while the effective refractive~index is $n_{\text{eff}} \!=\! 1.4$. The convergence threshold for the iterative algorithm is $\epsilon \!= \!10^{-4}$,  and $\rho=0$ dB. The transmit beamforming vector is initialized using maximum ratio transmission subject to the FD-BS transmit power constraint. Meanwhile, the transmit power of the UL-UE is initialized under its transmit power constraint. The TPAs and RPAs are initialized around their corresponding served users, and $\omega_{\text{DL}} \!=\! \omega_{\text{UL}} \!= \!1$. For performance evaluation, we consider two fixed antenna array benchmarks: 1) \textbf{Conv-$50 \text{ cm}$}, where the transmit and receive antenna arrays are centered in the service region with a spacing of $50$ cm \cite{ref16}; and 2) \textbf{Conv-$L$}, where the transmit and receive antennas are respectively centered at the two opposite edges separated by the length $L$ (i.e., at $x_{\text{T}} \!= \!-\frac{L}{2}$ and $x_{\text{R}} \!= \!\frac{L}{2}$).\looseness=-1
\begin{algorithm}
\caption{AO-based Algorithm for Solving Problem \textbf{P2}}
\label{Algorithm 1}
\begin{algorithmic}[1]
\STATE \textbf{Initialization:} Set iteration index $t$, initialize optimization variables and calculate initial objective value $U^{(0)}$.
\REPEAT
    \STATE Update $\mathbf{v}$,$u$ according to \eqref{eq:14} and \eqref{eq:15};
    \STATE Update MSE weights $\alpha, \beta$ according to \eqref{eq:13} ;
    \STATE Update $p_\text{t}$ according to \eqref{eq:17} and \eqref{eq:18};
    \STATE Update $\mathbf{w}$ according to \eqref{eq:21} and \eqref{eq:22};
    \STATE Update $\!\mathbf{x}^{\text{RPA}}\!$ and $\!\mathbf{x}^{\text{TPA}}\!$ by solving problem \eqref{eq:23};
\UNTIL $|U^{(t)} - U^{(t-1)}| / U^{(t-1)} < \epsilon$ or $t \ge T_{\text{max}}$;
\STATE \textbf{Output:} $\mathbf{v}^*, p_\text{t}^*, \mathbf{w}^*, \mathbf{x}^{\text{RPA}*}, \mathbf{x}^{\text{TPA}*}$.
\end{algorithmic}
\end{algorithm}

\begin{figure}[!t]
\vspace*{-0.20cm} 
\centering
\includegraphics[width=3.29in]{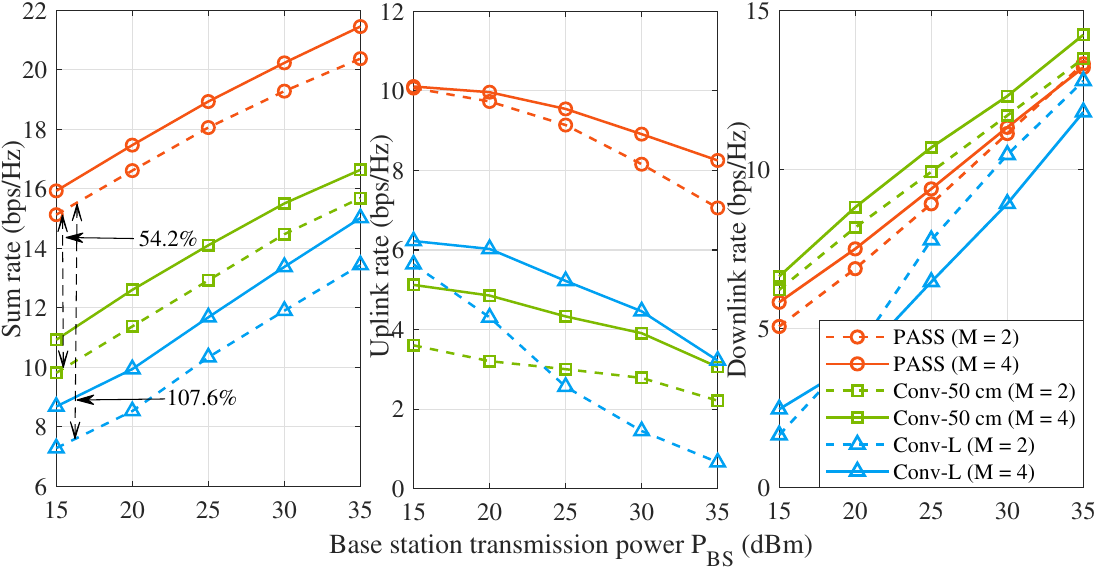}
\captionsetup{font=footnotesize, skip=2pt} 
\captionsetup{justification=raggedright, singlelinecheck=false}
\caption{User rates versus $P_{\text{BS}}$ with $K=1$, $P_\text{t}=15$ dBm.}
\label{Fig_2}
\end{figure}

\begin{figure}[!t]
\vspace{-0.8em}
\centering
\includegraphics[width=2.16in]{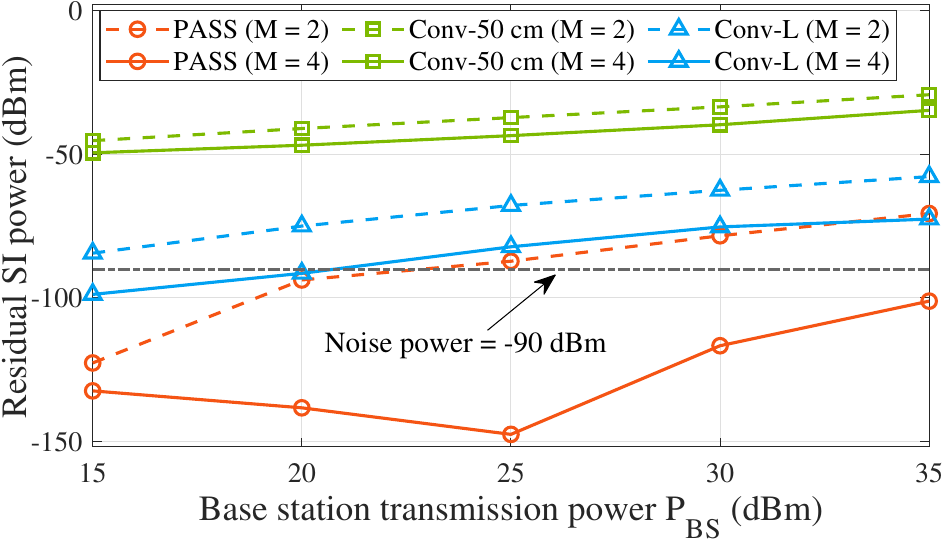}
\captionsetup{font=footnotesize, skip=2pt} 
\captionsetup{justification=raggedright, singlelinecheck=false}
\caption{{Residual SI power versus $P_{\text{BS}}$ with $K=1$, $P_\text{t}=15$ dBm.}}
\label{Fig_3}
\end{figure}

Fig. \ref{Fig_2} shows the user rates versus $P_{\text{BS}}$. As observed from the left subfigure, \textbf{PASS} significantly outperforms \textbf{Conv-$50 \text{ cm}$} and \textbf{Conv-$L$} in terms of the sum-rate. For example, when $M = 2$ and $P_{\text{BS}} = 15$ dBm, \textbf{PASS} achieves performance gains of 54.2\% and 107.6\% over \textbf{Conv-$50 \text{ cm}$} and \textbf{Conv-$L$}, respectively. Furthermore, for all considered schemes, as $P_{\text{BS}}$ increases, the uplink rate decreases, whereas the downlink rate increases. This is because a larger downlink power budget encourages the system to improve the downlink rate for sum-rate maximization, although the resulting stronger SI degrades the uplink transmission. 
To quantify the SI suppression capability of each scheme, Fig. \ref{Fig_3} plots the residual SI power before digital combining ($\|\mathbf{H}_{\text{SI}}\mathbf{w}\|^2$). As shown, \textbf{PASS} consistently maintains the lowest residual SI power in all configurations.

Since the receiver combining vector $\mathbf v$ performs SI~suppression in the digital-domain, its effectiveness is affected~by the signal quality prior to digital processing. In particular, the finite transceiver dynamic range may introduce additional distortion when strong SI is present and severely~degrade beamforming gains in multi-antenna architectures~\cite{ref1},~\cite{ref2},~\cite{ref14}. To consider this effect, we adopt a finite transceiver dynamic range model in the following evaluation. Following \cite{ref17}, we model the effect of the limited dynamic range by injecting independent zero-mean complex Gaussian distortion noise vectors $\mathbf{e}\! \sim \!\mathcal{CN}(\mathbf{0}, \mathbf{K}_{\text{RX}})$ and $\mathbf{c}\! \sim\! \mathcal{CN}(\mathbf{0}, \mathbf{K}_{\text{TX}})$ at each receive and transmit antenna of the FD-BS, respectively. The covariance matrices are defined as $\mathbf{K}_{\text{RX}} \!= \!\gamma\mathcal{D}(\mathbf{\Phi})$~and $\mathbf{K}_{\text{TX}}\! \!= \!\!\kappa\mathcal{D}(\mathbf{w}\mathbf{w}^H)$, where $\mathcal{D}(\mathbf{A})$ denotes a diagonal matrix containing the elements along the diagonal of $\mathbf{A}$, $\gamma, \kappa \!\ll \!1$ characterize the dynamic range of the receiver and transmitter, respectively, and $\mathbf{\Phi} \!\triangleq\! \text{Cov}\{\mathbf{y}_{\text{UL}}\}$ is the covariance matrix of the undistorted received signal at the FD-BS. Consequently, the SINR expressions under the limited dynamic range can be reformulated as
\begin{equation}
\label{eq:24}
\Gamma_{\text{DL}} = \frac{|\mathbf{h}_{\text{DL}}^\text{H} \mathbf{w}|^2}{|h_{\text{CCI}}|^2 p_\text{t} + \mathbf{h}_{\text{DL}}^\text{H} \mathbf{K}_{\text{TX}} \mathbf{h}_{\text{DL}} + \sigma_{\text{DL-UE}}^2},
\end{equation}
\begin{equation}
\label{eq:25}
\Gamma_{\text{UL}} = \frac{p_\text{t} |\mathbf{v}^\text{H} \mathbf{h}_{\text{UL}}|^2}{|\mathbf{v}^\text{H} \mathbf{H}_{\text{SI}} \mathbf{w}|^2 + \mathbf{v}^\text{H} \mathbf{R}_{\text{n}} \mathbf{v}},
\end{equation}
where $\mathbf{R}_{\text{n}} = \mathbf{H}_{\text{SI}} \mathbf{K}_{\text{TX}} \mathbf{H}_{\text{SI}}^\text{H} + \mathbf{K}_{\text{RX}} + \sigma_{\text{FD-BS}}^2 \mathbf{I}_{K}$.

%generates severe nonlinear distortion that significantly degrades the effectiveness of receive beamforming.

\begin{figure}[!t]
\vspace*{-0.21cm} 
\centering
\includegraphics[width=3.32in]{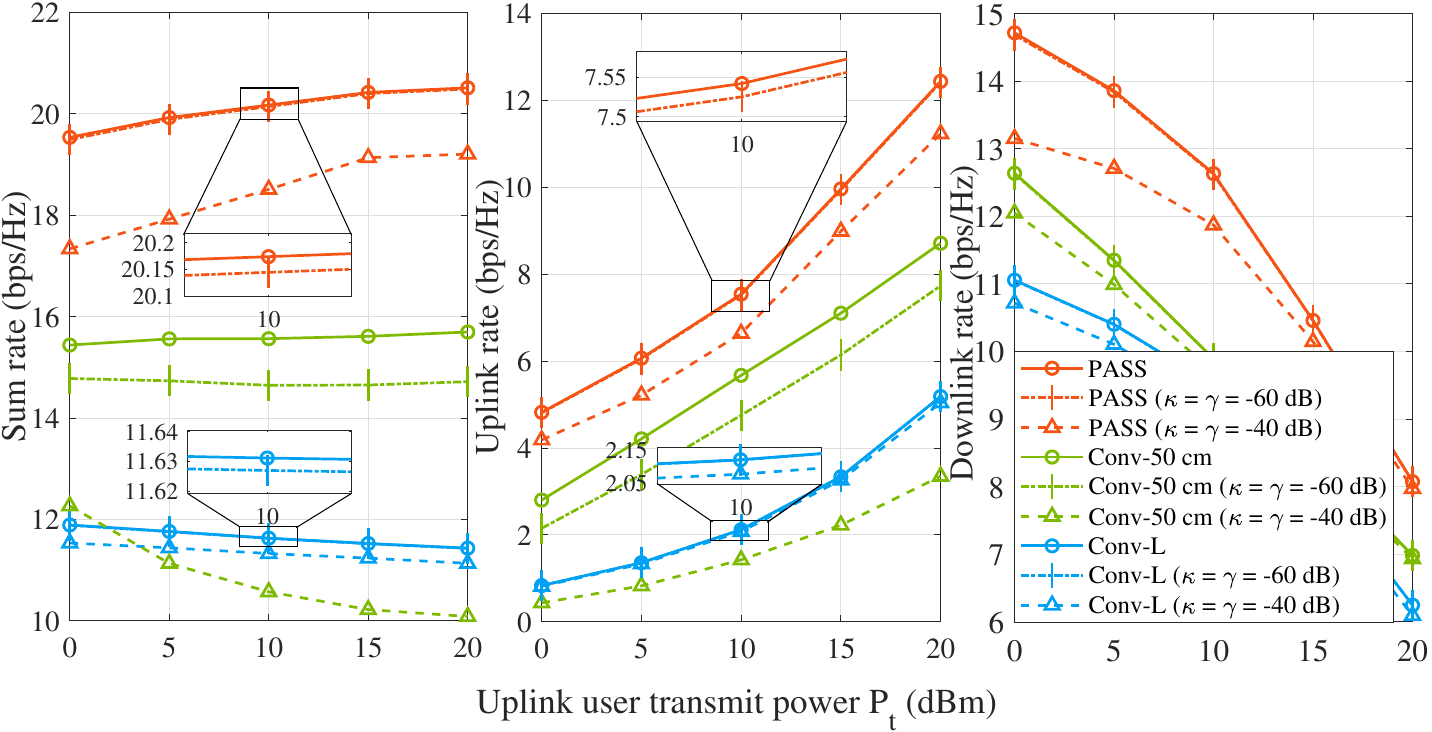}
\captionsetup{font=footnotesize, skip=2pt} 
\captionsetup{justification=raggedright, singlelinecheck=false}
\caption{User rates versus $P_\text{t}$ with $K=M=2$, $P_\text{BS}=30$ dBm.}
\label{Fig_4}
\end{figure}

Fig. \ref{Fig_4} shows the user rates versus $P_\text{t}$ under various dynamic range levels. As observed, higher $P_\text{t}$ improves uplink rates but inevitably degrades downlink rates due to exacerbated CCI. Among all schemes, \textbf{PASS} achieves the highest sum-rate as well as the best uplink and downlink performance. This superiority stems from its flexible antenna reconfiguration, which synergizes with transmit beamforming to effectively suppress SI in the spatial domain and enhance the desired~signal power, thereby mitigating the non-linear distortion caused by limited dynamic range. In contrast, under limited dynamic range conditions, the uplink rate of \textbf{Conv-$50 \text{ cm}$} experiences a non-negligible decrease. This~severe degradation occurs because its fixed architecture fails to adequately isolate strong SI, thereby inducing significant non-linear distortion that~degrades the effectiveness of receive beamforming. Although \textbf{Conv-$L$} is marginally~affected by the limited dynamic range as its large antenna separation attenuates SI before reception, its fixed deployment cannot~adapt to the user location, resulting in a lower~desired channel gain than \textbf{PASS}. \looseness=-1

\begin{figure}[!t]
\vspace{-0.6em}
\centering
\includegraphics[width=1.56in]{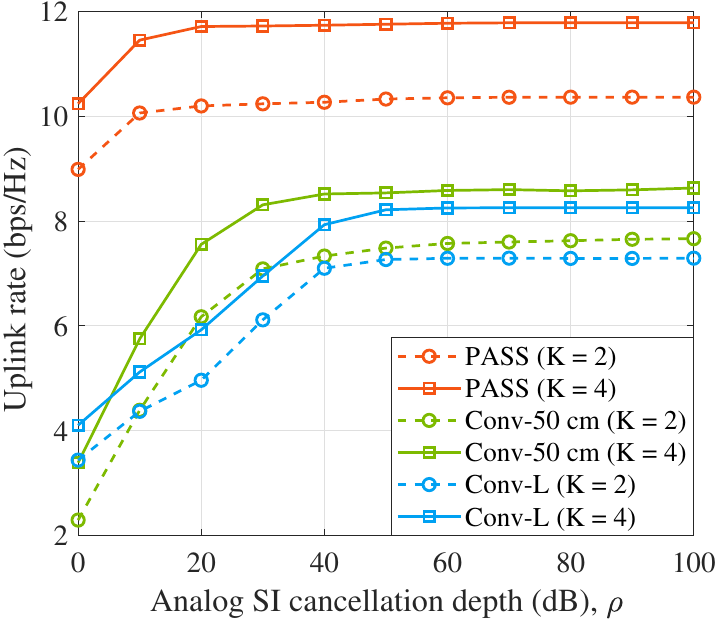}
\captionsetup{font=footnotesize, skip=2pt} 
\caption{Uplink rates versus analog SI cancellation depth with $M=2$, $P_\text{BS}=30$ dBm, $P_\text{t}=15$ dBm, and $\kappa = \gamma = -40$ dB.}
\label{Fig_5}
\end{figure}

Fig. \ref{Fig_5} illustrates the uplink rates versus the analog-domain SI cancellation depth $\rho$. As can be seen, both \textbf{Conv-$50 \text{ cm}$} and \textbf{Conv-$L$} reach their uplink saturated rates at an analog cancellation depth of approximately $40 \text{ dB}$. Although \textbf{Conv-$L$} outperforms \textbf{Conv-$50 \text{ cm}$} in the regime without analog cancellation via physical isolation, its saturated rate remains lower than those of \textbf{PASS} and \textbf{Conv-$50 \text{ cm}$} due to lower channel gains. In contrast, \textbf{PASS} achieves a substantial initial rate without additional cancellation and fully releases its uplink potential with merely about $10 \text{ dB}$ of analog cancellation. This advantage is attributed to the spatial channel reconfiguration capability of \textbf{PASS}, which simultaneously ensures the desired channel gain and SI suppression, effectively relaxing the stringent requirements for complex and costly analog SI cancellation in FD systems.
\vspace*{-0.35cm} 
\section{Conclusion}

This letter investigated a novel PASS-assisted FD system. We formulated a joint optimization problem to maximize the weighted sum-rate of the system. To tackle this highly non-convex problem, we reformulated it via the WMMSE framework and developed an efficient AO algorithm. Simulations verified that, benefiting from the large-scale reconfiguration of spatial channels, the PASS architecture ensures effective  simultaneous transmission for uplink and downlink while significantly improving the system sum-rate. In particular, it outperforms conventional fixed antenna arrays under both ideal and limited dynamic range conditions, substantially relaxing the stringent~requirements for costly and complex analog SI cancellation in FD systems.

\vspace*{-0.7cm} 
\bibliographystyle{IEEEtran}
\bibliography{ref}

% Generated by IEEEtran.bst, version: 1.14 (2015/08/26)
\begin{thebibliography}{10}
\providecommand{\url}[1]{#1}
\csname url@samestyle\endcsname
\providecommand{\newblock}{\relax}
\providecommand{\bibinfo}[2]{#2}
\providecommand{\BIBentrySTDinterwordspacing}{\spaceskip=0pt\relax}
\providecommand{\BIBentryALTinterwordstretchfactor}{4}
\providecommand{\BIBentryALTinterwordspacing}{\spaceskip=\fontdimen2\font plus
\BIBentryALTinterwordstretchfactor\fontdimen3\font minus
  \fontdimen4\font\relax}
\providecommand{\BIBforeignlanguage}[2]{{%
\expandafter\ifx\csname l@#1\endcsname\relax
\typeout{** WARNING: IEEEtran.bst: No hyphenation pattern has been}%
\typeout{** loaded for the language `#1'. Using the pattern for}%
\typeout{** the default language instead.}%
\else
\language=\csname l@#1\endcsname
\fi
#2}}
\providecommand{\BIBdecl}{\relax}
\BIBdecl

\bibitem{ref1}
B.~Smida \emph{et~al.}, ``Full-duplex wireless for {6G}: Progress brings new
  opportunities and challenges,'' \emph{IEEE J. Sel. Areas Commun.}, vol.~41,
  no.~9, pp. 2729--2750, Sep. 2023.

\bibitem{ref2}
B.~Smida \emph{et~al.}, ``In-band full-duplex: The physical layer,''
  \emph{Proc. IEEE.}, vol. 112, no.~5, pp. 433--462, May 2024.

\bibitem{ref3}
M.~Mohammadi \emph{et~al.}, ``A comprehensive survey on full-duplex
  communication: Current solutions, future trends, and open issues,''
  \emph{IEEE Commun. Surv. Tutorials.}, vol.~25, no.~4, pp. 2190--2244, 2023.

\bibitem{ref4}
T.~Fukui, K.~Komatsu, Y.~Miyaji, and H.~Uehara, ``Analog self-interference
  cancellation using auxiliary transmitter considering {IQ} imbalance and
  amplifier nonlinearity,'' \emph{IEEE Trans. Wireless Commun.}, vol.~19,
  no.~11, pp. 7439--7452, Nov. 2020.

\bibitem{ref5}
A.~Fukuda \emph{et~al.}, ``Pinching antenna: Using a dielectric waveguide as an
  antenna,'' \emph{NTT DOCOMO Tech. J.}, vol.~23, no.~3, pp. 5--12, 2022.

\bibitem{ref6}
Z.~Ding \emph{et~al.}, ``Flexible-antenna systems: A pinching-antenna
  perspective,'' \emph{IEEE Trans. Commun.}, vol.~73, no.~10, pp. 9236--9253,
  2025.

\bibitem{ref7}
M.~Sun, C.~Ouyang, S.~Wu, and Y.~Liu, ``Multiuser beamforming for
  pinching-antenna systems: An element-wise optimization framework,''
  \emph{IEEE Trans. Wireless Commun.}, vol.~25, pp. 6538--6552, 2026.

\bibitem{ref8}
A.~Bereyhi, C.~Ouyang, S.~Asaad, Z.~Ding, and H.~V. Poor, ``{MIMO-PASS}: Uplink
  and downlink transmission via {MIMO} pinching-antenna systems,'' \emph{IEEE
  Trans. Commun.}, vol.~74, pp. 5701--5716, 2026.

\bibitem{ref9}
S.~A. Tegos \emph{et~al.}, ``Minimum data rate maximization for uplink
  pinching-antenna systems,'' \emph{IEEE Wireless Commun. Lett.}, vol.~14,
  no.~5, pp. 1516--1520, May 2025.

\bibitem{ref10}
V.~E. Galanopoulou \emph{et~al.}, ``Viterbi state selection for discrete
  pinching antenna systems,'' \emph{IEEE Commun. Lett.}, vol.~30, pp.
  1523--1527, 2026.

\bibitem{ref11}
S.~Khisa \emph{et~al.}, ``Joint uplink and downlink resource allocation and
  antenna activation for pinching antenna systems,'' in \emph{Proc. IEEE
  Wireless Commun. Netw. Conf. (WCNC)}, Kuala Lumpur, Malaysia, 2026, pp. 1--6.

\bibitem{ref12}
C.~Ouyang \emph{et~al.}, ``Uplink and downlink communications in segmented
  waveguide-enabled pinching-antenna systems ({SWANs}),'' \emph{IEEE Trans.
  Commun.}, vol.~74, pp. 3688--3703, 2026.

\bibitem{ref13}
F.~Zhu, F.~Gao, T.~Zhang, K.~Sun, and M.~Yao, ``Physical-layer security for
  full duplex communications with self-interference mitigation,'' \emph{IEEE
  Trans. Wireless Commun.}, vol.~15, no.~1, pp. 329--340, Jan. 2016.

\bibitem{ref14}
M.~Wu, X.~Lei, and D.~B. da~Costa, ``{STAR-RIS}-assisted full-duplex symbiotic
  radio,'' \emph{IEEE Wireless Commun. Lett.}, vol.~13, no.~11, pp. 2945--2949,
  Nov. 2024.

\bibitem{ref15}
Q.~Shi \emph{et~al.}, ``An iteratively weighted {MMSE} approach to distributed
  sum-utility maximization for a {MIMO} interfering broadcast channel,''
  \emph{IEEE Trans. Signal Process.}, vol.~59, no.~9, pp. 4331--4340, Sep.
  2011.

\bibitem{ref16}
E.~Everett, A.~Sahai, and A.~Sabharwal, ``Passive self-interference suppression
  for full-duplex infrastructure nodes,'' \emph{IEEE Trans. Wireless Commun.},
  vol.~13, no.~2, pp. 680--694, Feb. 2014.

\bibitem{ref17}
B.~P. Day, A.~R. Margetts, D.~W. Bliss, and P.~Schniter, ``Full-duplex {MIMO}
  relaying: Achievable rates under limited dynamic range,'' \emph{IEEE J. Sel.
  Areas Commun.}, vol.~30, no.~8, pp. 1541--1553, Sep. 2012.

\end{thebibliography}
\end{spacing}
\end{document}